# Non-Hermitian heterostructure for two-parameter sensing

Jieun Yim[1], Han Zhao[2], Bikashkali Midya[1], Liang Feng[1]

[1]*Department of Materials Science and Engineering, University of Pennsylvania, Philadelphia, PA 19104, USA*
[2]*Department of Electrical and Systems Engineering, University of Pennsylvania, Philadelphia, PA 19104, USA*

Non-Hermitian systems at the exceptional point (EP) degeneracy are demonstrated to be highly sensitive to environmental perturbation. Here, we propose and theoretically investigate a novel multilayered heterostructure favoring double EPs for a unique set of material parameters at which forward- and backward-reflection coefficients vanish, respectively. Such an EP heterostructure is shown to scatter off light when system parameters are perturbed away from the degeneracies due to the effect of ambient temperature and mechanical stress fluctuations. The proposed structure is conducive in manipulating optical responses for two mutually independent parameters sensing.

The emerging field of non-Hermitian photonics offers deliberate control of light transport, detection, generation, and trapping in unprecedented ways, with revolutionizing possibilities for novel technological advancements [1-6]. Owing to the energy exchange between the environment, such open optical systems are often characterized by the existence of exceptional point (EP) degeneracies in their complex eigenvalue spectra. EPs are topological singularities in the parameter space where at least two eigenvalues and corresponding eigenstates coalesce simultaneously [7,8]. When a system, initially biased to an EP, is subject to an environment-mediated weak perturbation, the coalesced eigenvalues bifurcate. By measuring the resulting eigenvalues splitting, one can quantify the fluctuation in the surrounding environmental parameters which induce such splitting. This simple principle has triggered the recent theoretical and experimental developments of EP based nanophotonic sensors [9-13]. A key difference between EP and Hermitian degeneracy (e.g. a diabolic point) is that the former is "hypersensitive" to perturbations. In a system operating around a Hermitian degeneracy, the resulting eigenvalues splitting is linearly proportional to the perturbation strength, while in a non-Hermitian system the splitting scales as $N$-th root of a given perturbation strength, where $N$ denotes the order of an EP [10]. This results in an enhanced sensitivity of frequency splitting in non-Hermitian systems if the perturbation is very weak. Recent experiments have demonstrated hypersensitive optical sensing using engineered high-quality optical resonant cavities that operate near an EP [13,14].

The existence of an EP, however, is not limited to resonant systems. It can appear for the scattering states lying in the continuum as well. When incident from both forward and backward directions, scattering of optical waves by a one-dimensional non-Hermitian structure is characterized by a scattering matrix $S = \begin{pmatrix} t & r_b \\ r_f & t \end{pmatrix}$. Here, $t, r_f$ and $r_b$ denote transmission, forward- and backward-reflection coefficients, respectively. The scattering matrix has eigenvalues $\gamma_\pm = t \pm \sqrt{r_f r_b}$, and an EP degeneracy occurs when $r_f r_b = 0$ is satisfied.

Due to the unidirectional reflectionless condition inherited at an EP, only one of the $r_f$ or $r_b$ vanishes, but not both [15,16]. Therefore, perturbation of the corresponding vanishing reflection coefficient usually limits the detection of a "single" parameter. Indeed, in a recent experimental study, it has been demonstrated that an optical EP structure when judicially engineered in a microscopic glass slide can act as a thermal sensing imager in addition to the conventional tomographic imaging [17]. However, simultaneous detection of more than one environmental parameter by utilizing multiple exceptional points in a single structure has been an open problem— acquiring multiple EPs which satisfy a single set of system parameters is a rather challenging task. Nonetheless, the discovery of such a structure can be highly conducive for more versatile and multi-functional integrated sensor applications.

Here, we overcome this challenge partially by proposing a multi-layered, non-Hermitian, asymmetric structure, which embraces two 2nd order exceptional points for a unique set of system parameters. Lifting the EPs degeneracy simultaneously, by environmental perturbation, can facilitate to measure the fluctuations of two independent environmental parameters (such as temperature and stress). In order to demonstrate our sensing protocol, we consider a particular setup of a five layered refractive index profile, which can conveniently be integrated on a transparent glass substrate. As shown in Fig. 1(a), the structure consists of alternating layers of gold (Au) and two different widely-used polymers: low-density polyethylene (LDPE) and polymethyl methacrylate (PMMA). The structure is homogeneous in the *xy*-plane, but varies in the z-direction as

$$n(z) = \begin{cases} n_{\text{Au}}, & L_0 < z < L_1 \\ n_{\text{LDPE}}, & L_1 < z < L_2 \\ n_{\text{Au}}, & L_2 < z < L_3 \\ n_{\text{PMMA}} & L_3 < z < L_4 \\ n_{\text{Au}}, & L_4 < z < L_5 \end{cases} \qquad (1)$$

where $L_0 = 0$. The system is non-Hermitian because refractive index of Au is complex in the optical diapason. The polymers were selected due to their high linear thermal expansion coefficients and low elastic moduli (numerical values are given below). These properties render both the materials to deform due to the slight variations of temperature and stress. LDPE and PMMA are separately sandwiched between the Au layers such that the incident light in the forward direction is strongly affected by the deformation of LDPE, and the light in the backward direction is more influenced by that of PMMA.

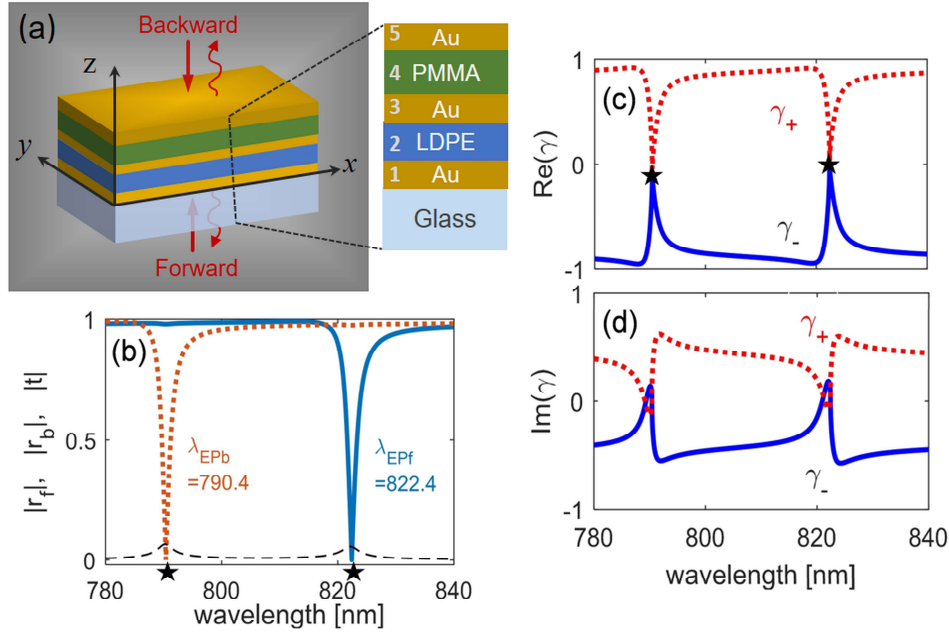

Fig. 1. **Sensing protocol.** (a) Schematic of the layered heterostructure. The material compositions of different layers are shown in the inset. The critical thickness of each layers for which the structure attains two EPs are given in the text. (b) Spectra of forward (blue-solid) and backward (orange-dotted) reflection and transmission (black-dashed). At the EPs, marked as '*' at $\lambda_{EPf}$ and $\lambda_{EPb}$, the forward and backward reflections become zero. (c) Real and (d) imaginary parts of the eigenvalue spectrum ($\gamma_\pm$) of the optical scattering matrix. Coalescence of $\gamma_\pm$ are seen at two critical values of wavelength which corresponds to two EPs. Any departure from EPs can be used for parameter estimation.

To theoretically predict the existence of double EPs, the reflection coefficients of this multilayer structure were calculated by the transfer matrix method [18]. The coefficients are given by

$$r_{f(b)} = \frac{M_{21}^{f(b)} + k_f k_b M_{12}^{f(b)} + i\left[k_{f(b)} M_{22}^{f(b)} - k_{b(f)} M_{11}^{f(b)}\right]}{-M_{21}^{f(b)} + k_f k_b M_{12}^{f(b)} + i\left[k_{f(b)} M_{22}^{f(b)} + k_{b(f)} M_{11}^{f(b)}\right]} \quad (2)$$

where $M_{ij}^{f(b)}$ are the elements of forward and backward transfer matrices $M^{f(b)} = M_{5(1)} M_{4(2)} M_{3(3)} M_{2(4)} M_{1(5)}$ with transfer matrix for the $j$-th layer

$$M_j = \begin{pmatrix} \cos(n_j k_0 \Delta L_j) & \frac{1}{n_j k_0}\sin(n_j k_0 \Delta L_j) \\ -n_j k_0 \sin(n_j k_0 \Delta L_j) & \cos(n_j k_0 \Delta L_j) \end{pmatrix}. \quad (3)$$

Here, $k_f = k_b = n_{air} k_0$, with $n_{air}=1$, and $k_0$ is the free space wavenumber. The refractive index and thickness of the $j$-th layer are given by $n_j$ and $\Delta L_j = (L_j - L_{j-1})$, respectively. We have assumed that the behavior of an EP is substrate independent. In the theoretical determination of the optimal parameters leading to the EPs, we, therefore, have ignored the substrate in the forward direction. Careful analysis, however, is required in order to consider the additional optical responses due to a substrate in realistic experiment. As mentioned earlier, the eigenvalues of scattering matrix become degenerate in the following two cases: $r_b \neq r_f = 0$ and $r_f \neq r_b = 0$. These constraints are fulfilled at different wavelengths $\lambda_{EPf} = 822.4$nm and $\lambda_{EPb} = 790.4$nm for forward and backward incidences, respectively; while the system parameters: $\Delta L_1=29.89$, $\Delta L_2=2395.09$, $\Delta L_3=59.8$, $\Delta L_4=2340.18$nm, $\Delta L_5=31.254$nm, $n_{LDPE}=1.51$, $n_{PMMA}=1.485$, remain valid for both cases at room temperature (~23°C). Note that the wavelength dependence of the refractive index for the Au layer: $n_{Au}(\lambda_{EPf}) = 0.162 + 5.0511\,i$ and $n_{Au}(\lambda_{EPb}) = 0.1499 + 4.8473\,i$. In the numerical computation of the reflection spectrum (as shown in Fig. 1b), we used the Drude model for varying complex refractive index of the Au layer: $n_{Au}(\omega) = \sqrt{1 - \omega_p^2/(\omega^2 + i\zeta\omega)}$, where $\omega_p = 1.881 \times 10^{15}$ Hz is the plasma frequency and $\zeta = 2.254 \times 10^{13}$ Hz is the damping coefficient. In Fig.1b, we have shown numerically calculated reflection spectra ($|r_f|$, and $|r_b|$) and the transmission coefficient ($|t|$) versus the wavelength. The spectrum shows the unidirectional reflectionlessness at the two critical wavelengths $\lambda_{EPf}$ and $\lambda_{EPb}$ for which forward and backward reflection vanishes, respectively. The real and imaginary parts of scattering matrix eigenvalues [shown in Fig. 1(c) and (d)] confirm that such critical points correspond to the EP degeneracies. At these EPs, the transmission is found to be negligibly small because of the strong absorption by the Au layers.

When the system is designed to operate around the double EPs, small environmental perturbation may deform the material properties. This in turn lifts the EPs degeneracy, making the system to respond to the incident light at the critical wavelengths. The basic sensing mechanism of this optical structure is thus based on the fact that once EP degeneracy is lifted upon perturbations, the reflections at $\lambda_{EPf}$ and $\lambda_{EPb}$ are no longer zero. In our multilayer system, the deviation from EP can inevitably come from the surrounding environmental fluctuation. In the following we have considered the effect of temperature change ($\delta T$) and mechanical stress ($\delta \sigma$) on the reflection spectra of the heterostructure. The transmission coefficient plays no role in the sensor operation. Since polymers are subject to thermo-optic effect and thermal expansion when there is a temperature change, and also easily experience strain under mechanical stress, the deviation from EP occurs when the thicknesses and the refractive indices of the polymers change. Due to the perturbation, the refractive index ($n_j$) and the thickness ($\Delta L_j$) of the LDPE and PMMA layers change according to: $n_j \to (n_j - \beta_j \delta T)$ and $\Delta L_j \to (1 + \alpha_j \delta T)(1 - \delta\sigma/E_j)\Delta L_j$, where $\beta_j$, $\alpha_j$ and $E_j$ are the thermo-optic coefficient, thermal expansion linear coefficient and elastic modulus of the layers, given by $\beta_{LDPE}$: $3.5 \times 10^{-4}$ °C$^{-1}$ [19], $\beta_{PMMA}$: $1.37 \times 10^{-4}$ °C$^{-1}$ [20], $\alpha_{LDPE}$: $18 \times 10^{-5}$ °C$^{-1}$ [21], $\alpha_{PMMA}$: $7.3 \times 10^{-5}$ °C$^{-1}$ [22], and $E_{LDPE}$: 282 MPa [21], $E_{PMMA}$: 3240 MPa [21]. Note that the temperature dependence of $E_j$ is neglected for sufficiently small temperature change (in the following we have considered up to $\delta T$=10°C above the room temperature). Owing to the inadequate experimental data, we have also neglected the stress-optic effect on the refractive index change of the polymer layers. In the presence of such an effect, the qualitative characteristics of the optical responses, as shown in Fig. 2, are expected to be unchanged.

In order to gain insight about the effect of the change of temperature and stress on the reflection, we analytically approximate the reflection coefficients by considering only the first order terms of various small quantities in the perturbed transfer matrices $M_2$ and $M_4$ corresponding to the LDPE and PMMA layers. To do this aim, we have approximated the perturbation in the argument appeared in Eq.(3): $n_j k_0 \Delta L_j \approx n_j k_0 \Delta L_j + \epsilon_j$, with

$$|\epsilon_j| = |k_0 \Delta L_j [(\alpha_j n_j - \beta_j)\delta T - n_j \delta\sigma/E_j]| \ll 1, \quad (4)$$

$j$=2,4, and used the fact that $\cos \epsilon_j \sim 1$ and $\sin \epsilon_j \sim \epsilon_j$, in the analytical derivation of the reflection coefficients given below. Note here that the quantities $\delta T$ and $\delta\sigma$ need not be small in order to have the above "smallness" to be valid. This benefits the consideration of a wide range of temperature and stress variations to take into account in the theory. Consequently, the relation between the forward and backward reflection coefficients $r_{f(b)}$ as a function of $\delta T$ and $\delta\sigma$, when the system is deviated from their corresponding EPs due to perturbation, have been derived after a straight-forward calculation as

$$r_{f(b)}(\delta T, \delta\sigma) \approx \frac{A_{f(b)} \delta T + B_{f(b)} \delta\sigma}{C_{f(b)} + D_{f(b)} \delta T + G_{f(b)} \delta\sigma} \quad (5)$$

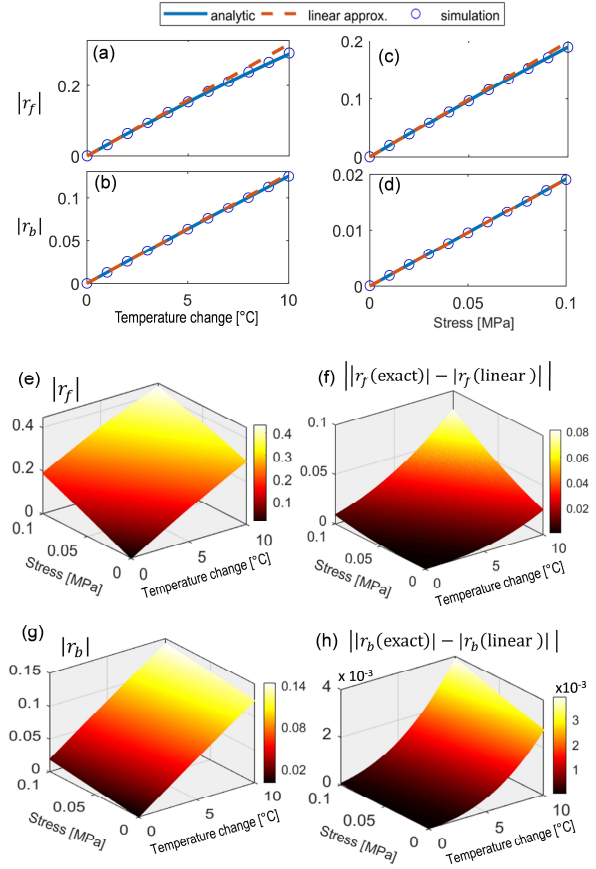

Fig. 2. **Parameters estimation.** (a) Numerically simulated, analytical non-linear and linear approximations of reflection spectra are shown when the heterostructure is deformed due to: (a), (b) temperature changes but stress remains zero, and (c), (d) applied stress but temperature changes is considered to be zero. (e) and (g) show forward and backward reflections when temperature and stress changes simultaneously. (f) and (h) show the absolute error between the linear approximation [Eq. (6)] and the exact results. In all these figures the temperature change is considered up to 0-10°C increment above room temperature, while stress is considered up to 0-0.1MPa.

where $A$, $B$, $C$, $D$ and $G$ are independent of $\delta T$ and $\delta\sigma$. The expressions of these quantities are too involved to be detailed here. $r_f$ and $r_b$ in Eq. (5) are calculated at the fixed wavelengths $\lambda_{EPf}$ and $\lambda_{EPb}$, respectively. Eq. (5) analytically describes how non-zero reflections arise upon temperature and stress change. In the limit of $\delta T = \delta\sigma$=0, we have both $r_f = 0$ and $r_b = 0$ which correspond to the forward and backward reflections vanishing at the double EPs, respectively. Note the nonlinear dependence of the $r_{f(b)}$ on the temperature and stress change in equation (5). The resultant analytical expressions of $|r_f|$ and $|r_b|$ are plotted against temperature change and stress separately in Fig. 2(a), (b), (c), and (d) in solid lines. The forward and backward reflections differ drastically with respect to the same amount temperature (or stress) change because of the asymmetry of the heterostructure. In order to confirm the effectiveness of the derived analytical expressions, the reflection coefficients are calculated by numerical simulation without approximations. These numerical results are also shown against temperature change and stress in the same figures (in "circles"). Our analytically approximations matched fairly accurately with those of simulated results. In Fig. 2(e) and (g), we have shown the full panorama of forward and backward reflections versus simultaneous variations of both the temperature and stress. From the figures, it is obvious that with increasing temperature change or applied stress, the system starts to deviate from the EP, and the reflection increases more as the system is farther away from the EPs.

For given measured values of $r_f$ and $r_b$, the change in the temperature and stress can be estimated by inverting the two equations, given in (5), to obtain $\delta T(r_f, r_b)$ and $\delta\sigma(r_f, r_b)$. It is straightforward to verify the nonlinear dependence of both $\delta T$ and $\delta\sigma$ on $r_f$ and $r_b$. In realistic experiments, however, it is desirable to match the measurement results with a linear dependence. Therefore, we have approximated the Eq. (5) further, by neglecting the terms nonlinear in small quantities, such that

$$\begin{pmatrix} r_f \\ r_b \end{pmatrix} \approx \begin{pmatrix} A_f/C_f & B_f/C_f \\ A_b/C_b & B_b/C_b \end{pmatrix} \begin{pmatrix} \delta T \\ \delta\sigma \end{pmatrix}. \qquad (6)$$

This linear sensitivity matrix simplifies the relation between the reflection and the multi parameters. As shown in Fig. 2, the above linear approximation predicts the actual reflection spectrum quite well when perturbation is small i.e. close to the EPs. The deviation of linear model from the exact values is apparent for forward reflections [as shown in Fig. 2(f)], when both the effect of temperature and stress change are taken into account. The error is relatively small for backward reflections [Fig. 2(h)]. The linear model is thus applicable in a small range of parameters close to the EPs. The neglected higher order terms will affect the accuracy with an overall larger perturbation.

Finally, we comment on the practical implementation of our proposed heterostructure in the state-of-the-art sensing devices. Although, the structure is few microns in thickness, the other two dimensions are considered to be arbitrary in the theory. This allows to install the heterostructure on a glass substrate, and the optical responses due to reflection of forward and backward incidences from two lasers, operating at $\lambda_{\text{EPf}}$ and $\lambda_{\text{EPb}}$, can be registered by photodetectors. The application of the compressive stress and temperature change on the sample can judiciously be performed, for example, by a glass-tip attached to a force transducer and by a thermo-electric device, respectively. The applied stress may additionally affect the temperature characteristics of the sensor, which requires further analysis. The sensitivity of our sensor can be probed in two ways. First, by the individual measurement of the forward and backward reflection coefficients, both of which vary linearly with respect to the temperature and stress variations in the vicinity of EP [Eq. (6)]. In this case, the optical responses (change in the reflection coefficients) are $\Delta|r_f| = 0.031$ and $\Delta|r_b| = 0.013$ for 1°C temperature change from the EP condition. Whereas, these values are 0.019 and 0.0018 for 10kPa stress change from the EP condition. The second method of probing the sensitivity is by measuring the eigenvalues splitting, $\Delta\gamma/2 = \left|\sqrt{r_f r_b}\right|$, at $\lambda_{\text{EPf}}$ and $\lambda_{\text{EPb}}$ respectively [17]. The latter offers much larger optical responses and thus higher sensitivity. For example, $\Delta\gamma(\lambda_{\text{EPf}})/2 = 0.158$ and $\Delta\gamma(\lambda_{\text{EPb}})/2 = 0.103$ for 1°C temperature change, whereas these values are 0.121 and 0.0347, respectively, for 10kPa stress change from the EP condition.

To summarize, we have proposed a two-parameters sensing protocol based on a non-Hermitian optical heterostructure favoring double EPs. The EPs satisfy a unique set of material parameters at which both the forward and backward reflections are zero. The zero-reflections are obtained at two different critical wavelengths $\lambda_{\text{EPf}}$ and $\lambda_{\text{EPb}}$, respectively. When the system parameters deviate from the EPs condition, for example due to the ambient temperature and stress fluctuations, the structure is shown to be sensitive and can reflect light in either direction. By measurement of the reflection spectrum at two critical wavelengths, one can estimate the fluctuations in the environmental parameters. The EP optical structures can, thus, be qualified as highly-efficient low-noise detectors for two independent parameters. The optical sensors enable the measurement of non-optical parameters by simply observing the optical spectrum without direct measurement tools such as a thermometer, a pressure gauge and other means. It is worth mentioning here that the simultaneous sensing of both temperature and stress are studied earlier in fiber Bragg grating Hermitian structures [23]. However, because of the non-Hermitian EP, our model is hypersensitive and can easily be used in integrated nano-optics. The generalization of the proposed scheme to more than two parameters sensing applications using multi-mode scattering technique is a future direction of research.

*Funding.* National Science Foundation (ECCS-1811393, CBET 1706050, IIP-1718177); King Abdullah University of Science & Technology (OSR-2016-CRG5-2950-04); Army Research Office (ARO) (W911NF-15-1-0152).